\documentclass[11pt]{article}
\usepackage[a4paper, total={6in, 8in}]{geometry}
\usepackage{lipsum,bm}
\usepackage{amssymb}

\usepackage{mathtools}
\usepackage{caption,comment,verbatim}
\usepackage{subcaption}
\usepackage{float}
\usepackage{amsmath}
\usepackage{lastpage}
\usepackage{fancyhdr}
\usepackage{color,url}

\usepackage{commath}
\usepackage{blindtext}

\title{Modelling intermediate internal waves with currents and variable bottom }

\author{Rossen Ivanov $^{a,\dagger,}$\thanks{member of the Institute for Advanced Physical Studies, 111 Tsarigradsko shose Blvd., Sofia 1784, Bulgaria}  \phantom{8} and Lyudmila Ivanova $^{\ddagger}$
\phantom{*}
\\ School of Mathematics and Statistics, \\Technological University Dublin, \\ City Campus, Grangegorman Lower, \\ Dublin D07 ADY7, Ireland\\
\phantom{8}\\
$^a$ Email: Rossen.Ivanov@tudublin.ie \\
$^{\dagger}$ \url{https://orcid.org/0000-0003-1008-8272} \\
$^{\ddagger}$ \url{https://orcid.org/0009-0005-3984-8679}
 }

\begin{document}

\maketitle

\begin{abstract}
A model for internal interfacial waves between two layers of fluid in the presence of current and variable bottom is studied in the flat-surface approximation. Fluids are assumed to be incompressible and inviscid. Another assumption is that the upper layer is considerably deeper with a lower density than the lower layer. The fluid dynamics is presented in Hamiltonian form with appropriate Dirichlet-Neumann operators for the two fluid domains, and the depth-dependent current is taken into account.   
The well known integrable Intermediate Long Wave Equation (ILWE) is derived as an asymptotic internal waves model in the case of flat bottom. For a non-flat bottom the ILWE is with variable coefficients. Two limits of the ILWE lead to the integrable Benjamin–Ono and Korteweg-de Vries equations. Higher-order ILWE is obtained as well.
\end{abstract}

\section{Introduction}
As a first approximation in oceanography, the ocean can be modeled as two large superimposed layers: the upper layer of warmer water and the lower layer of cold water of higher salinity. The boundary between the two layers is known as the {\it thermocline}. Thermocline is a very thin layer that separates the two large layers and where the temperature changes rapidly from its high value at the upper layer to its low value at the lower layer \cite{Boyd}. Since the density is a function of the temperature and salinity, the two layers also have different densities, the lower layer having a higher density. A further approximation can be made that the density is constant in each layer. The internal gravity waves form on the interface between the two layers, that is, on the thermocline. The depth of the thermocline is most often defined as the depth where the vertical temperature gradient is greatest. The thermocline has a mean depth of about 75-100 m along the equator; however, it varies in depth.  During summer, the thermoclines move deeper as the sun warms up a larger amount of water near the surface. In winter, on the other hand,  the thermocline depths decrease as the surface waters become colder \cite{Bow}. Therefore, in principle, we can have a large interval of values for the ratio of the depths of the two main layers.

The motion of the thermocline is also affected by currents. Currents are steady mean flows of ocean water in a prevailing direction, while waves are periodic motion of water disturbances. Currents are mostly related to mass transport, which involves mainly water and salt, but also pollutants and nutrients. For example, the major surface flow is the Equatorial Undercurrent (EUC), it is centered on the geographical equator. The EUC is driven by the wind and achieves velocities of up to 1.5 m/s \cite{Boyd}.  

The wave and current components of the velocity field (the vector field, which describes the velocity of fluid particles at a given point in space and time) interact in a complicated way due to the non-linear nature of the equations of ocean dynamics \cite{Constantin_2011,CJ}. Assuming a two-dimensional simplified picture of the fluid motion, in the case of current with a linear depth dependence, the current introduces only a constant vorticity. This fact significantly simplifies any further considerations, since the vorticity equation (that is, the equation for the time-evolution of the vorticity) is automatically satisfied.  The Hamiltonian formulation of the irrotational case \cite{CGK} allows for an extension that accommodates currents with piecewise linear shear \cite{Compelli1Wavemotion,CompelliIvanov1,CoIv2,CIM-16,CI-19,CuIv,CuIv2,Iv17}.

Our aim is to model intermediate long internal waves (when the thermocline is closer to the bottom). This propagation regime can be determined by the magnitudes of the typical wavelength $L$ and the depths of the fluid layers. In particular, if the upper fluid layer is of depth $h_1,$ (much deeper than the lower layer), then intermediate long waves are observed when $0.05 < h_1/L < 0.5.$
We consider a variable bottom, with some assumptions for the nature of the bottom variations, explicitly specified in Section \ref{sect2}.

In Sections \ref{sect2} and \ref{sect3} we introduce the problem setup and formulate the problem mathematically. Then in Section \ref{sect4} the problem is reformulated as a quasi-Hamiltonian system for two physical variables. The physical scales of the intermediate long-wave model and all other assumptions are introduced explicitly in Section \ref{sect5}. The one-component Intermediate Long Wave Equation (ILWE) is derived in Section \ref{sect6} and its limits, giving the Korteweg-de Vries (KdV) equation and the Benjamin-Ono (BO) equations, are obtained in Section \ref{sect7}. The effects of the variable bottom are analysed in Section \ref{sect8}. The higher order ILWE is derived in Appendix 1.

\section{Preliminaries - Setup and governing equations}\label{sect2}

\begin{figure}[!ht]
\centering
\includegraphics[width=0.6 \textwidth]{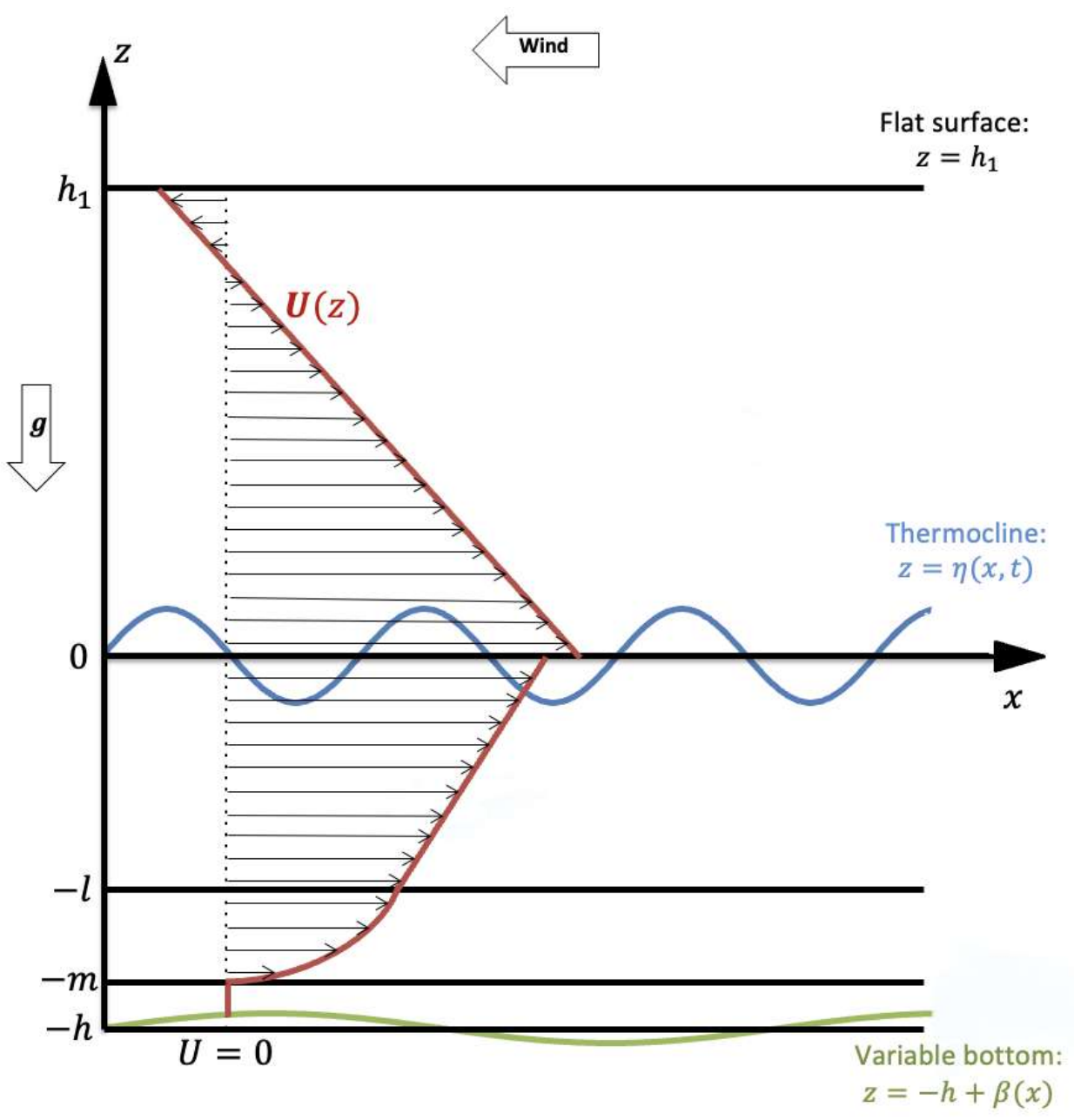}
 \caption{Coordinates and sketch of the fluid domain. The flat surface is given by $z=h_1,$ the free interface of the internal wave is $z=\eta(x,t)$ and is shown by the blue curve, the variable bottom is $z=-h+\beta(x)$ and is shown by the green curve. The current profile is schematically shown in red, its magnitude $\bm{U}(z)$ is accommodated in the horizontal direction. }\label{fig1}
\end{figure}

We consider an interfacial internal wave, represented by the surface $z=\eta(x,t)$, the thermocline, which separates the lower layer of fluid with higher constant density $\rho$ in the domain  
\begin{equation} \label{botl} \Omega(\eta, \beta):=\{(x,z), \,\, -h+\beta(x)<z<\eta(x,t)\},\end{equation} and the upper layer of fluid with lower constant density $\rho_1$ in the domain \begin{equation}\label{topl}\Omega_1(\eta):=\{(x,z) ,\,\,  \eta(x,t)<z<h_1\},\end{equation}
where $\beta(x)$ is the function, modeling the variations of the bottom, while $h$ and $h_1$ are the average depths of the two layers, that is, $z=h_1$ is the location of the flat surface, and $z=-h$ is the average depth of the bottom, cf. Fig. \ref{fig1}, mathematically this is expressed as
 \begin{equation}\label{eq:level}
     \int_{\mathbb{R}}\eta(x,t)dx=0, \qquad \int_{\mathbb{R}}\beta(x)dx=0.\end{equation} 
 The space-average value of a quantity $\mathfrak{h}(x, t)$ is the ratio of $\int_{\mathbb{R}}\mathfrak{h}(x,t)dx$ and the length of the space interval $\int_{\mathbb{R}} dx$ (which is infinity, the interval is the whole real axis). Thus, the space-average values of $\eta$ and $\beta$ will still be zero if the integrals \eqref{eq:level} result in finite quantities. As a rule, all quantities associated with the domain $\Omega_1$ will have a subindex 1. 
 
In addition, for simplicity, we assume some properties for the functions $\eta(x,t), \beta(x)$ that guarantee the existence of certain integrals. We assume that these functions belong to the Schwartz space of functions $\mathcal{S}(\mathbb{R})$ with respect to the $x$ variable (for any value of $t$). This guarantees a rapid decay of the functions with $|x|$ and reflects the localised nature of the wave disturbances and bottom variations.  In particular, this assumption implies 
 $$\lim_{x\to \pm \infty}\eta(x,t)=0, \qquad \lim_{x \to \pm \infty}\beta(x)=0.$$  
The function $\bm{U}(z)$ represents the background shear current. The current profile is piecewise linear with respect to $z$ with the exception of the layer near the bottom, where it gradually decays to zero. More specifically, the current profile is given by the function:
\begin{equation} \label{U5}
\bm{U}(z)=\left\{\begin{array}{lll}\gamma_1 z+\kappa_1, &\rm{when} & \eta(x,t) \le z \le h_1,\\
\gamma z + \kappa, & \rm{when} & -l\le z \le \eta(x,t), \\
U_{lm}(z) , &\rm{when} & -m  \le z \le -l, \\
0, &\rm{when} & -h +\beta(x) \le z \le -m.\end{array}\right.
\end{equation} Here  $\gamma$ and $\gamma_1$ are the constant vorticities in the fluid domains $\Omega$ and $\Omega_1$ respectively, $\kappa$ and $\kappa_1$ represent the constant component of the current at each layer, if $\kappa=\kappa_1$ the function $\bm{U}(z)$ is continuous at $z=0,$ that is, there is no slip between the two layers at $z=0$. For our considerations we assume this condition.  The illustration on Fig. \ref{fig1} corresponds to a situation where the current changes in the East-West direction and the winds are blowing to the west, for example, this is the situation with the Equatorial undercurrent, \cite{CI-19,CIM-16,CompelliIvanov1,CJ} so that the current on the surface is in the direction of the wind.  The continuity of the current in the presence of waves deserves special attention. 
The current is not continuous at $z=\eta$, that is, when $\eta\ne 0$ the current has a finite jump (continuous is only the normal component of the velocity field on $z=\eta(x,t)$). The layer between depths $z=-m$ and $z=-l$ is assumed to have current with profile $U_{lm}(z)$ in a general form (which could be chosen, for example, as necessary to match the field data). We choose $U_{lm}(z)$ as a continuous function such that $U_{lm}(-m)=0,$ $U_{lm}(-l)=-l\gamma+\kappa.$  As shown in \cite{CompelliIvanov1}, the energy per unit length of this layer is constant and thus the layer does not contribute to the equations at the interface (the thermocline). However, the reconstruction of the velocity field in the fluid body depends on the function $U_{lm}(z).$  The current gradually attenuates with depth and near the bottom layer it is zero, that is, there the motion is irrotational. The current profile is shown illustratively in Fig. \ref{fig1} in red for the situation of an undisturbed fluid when $\eta\equiv 0.$ 

We introduce the following notation for the velocity field in the two fluid domains as follows. For the horizontal component,
\begin{equation}
\bm{u}(x,z,t):=
\left\{\begin{array}{ccc}  u(x,z,t), &{\rm in} & \Omega,\\ u_1(x,z,t), &{\rm in} &\Omega_1,\end{array}\right.
\end{equation}
and for the vertical component
\begin{equation}
\bm{w}(x,z,t):=
\left\{\begin{array}{ccc} w(x,z,t), &{\rm in} & \Omega,\\ w_1(x,z,t), &{\rm in} &\Omega_1.\end{array}\right.
\end{equation}
In addition, we also assume incompressibility, which gives the equation of mass conservation 
\begin{equation}\label{masscons}
{ u}_x+{w}_z=0\,\,\textrm{in}\,\,\Omega , \qquad  { u}_{1,x}+{w}_{1,z}=0\,\,\textrm{in}\,\,\Omega_1.
\end{equation}
The mass conservation \eqref{masscons} ensures the existence of a stream function, $\psi(x,z,t)$  in $\Omega$ and $\psi_1(x,z,t)$  in $\Omega_1,$ determined up to an additive term that depends only on time, by
\begin{equation}\label{stream_func}
\left\{
\begin{array}{lll}
 u=\psi_z, & w=-\psi_x, & {\rm in} \quad \Omega,\\
u_1=\psi_{1,z}, & w_1=-\psi_{1,x}, & {\rm in}\quad\Omega_1
\end{array}\right.
\end{equation}
Fluid kinematics does not allow the transverse motion of particles through the fluid interface $z=\eta(x,t)$, that is, the normal velocity components are equal across the interface $z=\eta(x,t)$, leading to the so-called kinematic boundary condition on the interface,
\begin{equation}\label{KBC_gen}
    \eta_t=(w_{1})_{s}-\eta_x( u_1)_s = (w)_{s}- \eta_x (u)_s
\end{equation} where the subindex $s$ means that the evaluation is on $z=\eta(x,t).$
The condition \eqref{KBC_gen} means that the stream function is continuous on the interface $z=\eta(x,t)$, indeed, with \eqref{stream_func} and \eqref{KBC_gen} we obtain 
\begin{equation}
 \psi(x,\eta(x,t),t)=\psi_1(x,\eta(x,t),t)=-\int_{-\infty}^x \eta_t(x',t) dx'= - \partial_x ^{-1} \eta_t.
\end{equation} We also use the notation $$\chi(x,t):= (\psi)_s=(\psi_1)_s= - \partial_x ^{-1} \eta_t.$$
There are also kinematic conditions on the flat surface, \begin{equation}
 \label{kin_fs}
 w_1=0\,\,{\rm on}\,\,z=h_1,
\end{equation}
and on the variable bottom, which ensures that the normal velocity component at the bottom is zero, 
\begin{equation}
\label{KBC_VarBotta}
(w )_b -\beta_x (u) _b=0,
\end{equation}
where the subindex $b$ means that the evaluation is at the bottom, $z=-h + \beta(x)$.

One of the main assumptions of our analysis is about the structure of the velocity field. We explicitly separate the current by 
introducing a (generalized) velocity potential $$\bm{\varphi}=\left\{\begin{array}{ccc} \varphi, &{\rm in}& \Omega,\\ \varphi_1, &{\rm in} & \Omega_1,\end{array}\right.$$  such that 
\begin{equation}\label{vel_pot}
   \bm{u}=\bm{\varphi}_x+ \bm{U}(z) , \qquad \bm{w}=\bm{\varphi}_z.
\end{equation} The functions $\varphi(x,z,t)$ and $\varphi_1(x,z,t)$ are assumed in $\mathcal{S}(\mathbb{R})$ with respect to $x$ for any appropriate values of $z$ and all values of $t.$
The kinematic boundary conditions  \eqref{KBC_gen}, \eqref{kin_fs} and \eqref{KBC_VarBotta} can now be written as
\begin{equation}\label{kin_surface}
  \varphi_{1,z}=0 \, \, \text{on the surface} \, \, {z=h_1};
\end{equation} 
\begin{equation}\label{kin_interface}
 \begin{array}{c}
   \eta_t=( \varphi_{1,z})_{s}-\eta_x [(\varphi_{1,x})_{s}+\gamma_1\eta+\kappa],\\
 \eta_t=(\varphi_z)_{s}- \eta_x [(\varphi_x)_{s}+\gamma\eta+\kappa]
  \end{array}
\end{equation} on the interface and, 
\begin{equation}
\label{KBC_VarBott}
(\varphi_z)_b=\beta_x(\varphi_x)_b,
\end{equation}
on the bottom, since there is no current at the bottom, and the velocity field there is irrotational.

\section{Bernoulli's equation}\label{sect3}

The fluid dynamics is described by Euler’s equations 
\begin{equation}
 \left\{\begin{array}{lcl}
        \bm{ u}_t+\bm{u}\bm{u}_x+\bm{w}\bm{u}_z &=& -\bm{\frac{1}{\rho}}P_x,\\
         \bm{w}_t+\bm{u}\bm{w}_x+\bm{w}\bm{w}_z &=& -\bm{\frac{1}{\rho}}P_z-g,
        \end{array}\right.
\end{equation}
where $P=P(x,z,t)$ denotes the pressure, $g$ is the gravitational acceleration and $\bm{\rho}$ denotes the piecewise constant 
fluid density,  
\begin{equation} 
\bm{\rho}:=
\left\{\begin{array}{ccc} \rho, &{\rm in} & \Omega,\\ \rho_1 , &{\rm in} &\Omega_1,\end{array}\right.
\end{equation} where $\rho>\rho_1$ is assumed for the stability of the fluid system.

Euler's equations can be expressed by means of the stream function and of the generalized velocity potential in each fluid domain. These expressions are
$$\varphi_{1,t}+\frac{1}{2}|\nabla\psi_1|^2 -\gamma_1 \psi_1 +\frac{P_1}{\rho_1}+gz=f_1(t)\quad{\rm in}\quad \Omega_1,$$
$$\varphi_{t}+\frac{1}{2}|\nabla\psi|^2 - \gamma \psi +\frac{P}{\rho}+gz=f(t)\quad{\rm for}\quad -l\le z \le \eta(x,t).$$
On the top surface $P=P_{atm},$ the atmospheric pressure, hence $P_{atm}/\rho_1+gh_1=f_1(t)= $ const (at $x \to \pm \infty$ the other terms vanish). 
At the interface the pressures of the two layers are equal, that is, $P(x,\eta)=P_1(x,\eta)$ at $z=\eta(x,t).$ Thus, in a simmilar manner we obtain $f(t)=\rho_1 f_1(t)/\rho$  and the Bernoulli type equation
\begin{equation}
 \label{Bint}
 \rho\left[(\varphi_t)_s+\frac{|\nabla\psi|_s^2}{2}- \gamma \chi+g\eta\right]=\rho_1\left[(\varphi_{1,t})_s+\frac{|\nabla\psi_1|_s^2}{2}- \gamma_1 \chi+g\eta\right],
\end{equation}
or,
\begin{equation}
 \label{Ber}
 ((\rho\varphi -\rho_1 \varphi_1)_t)_s+\frac{\rho |\nabla\psi|_s^2 -\rho_1 |\nabla\psi_1|_s^2 }{2}- (\rho \gamma - \rho_1 \gamma_1)\chi+g(\rho-\rho_1)\eta = 0. 
\end{equation}
The Bernoulli equation \eqref{Ber} describes the time evolution of the velocity potentials. More precisely, it suggests that there is a particular variable $\xi:=(\rho\varphi -\rho_1 \varphi_1)_s,$ that describes fluid dynamics, together with $\eta(x,t),$ that evolves according to \eqref{kin_interface}. It turns out that there is a closed system of equations for the variables $\xi$ and $\eta.$ This system can be written in a convenient quasi-Hamiltonian form through the use of the so-called Dirichlet-Neumann operators \cite{CGK,CGS,Lan08}. Moreover, the difference between the flat and variable bottom cases is only in the expressions for the depth-dependent Dirichlet-Neumann operators. We refer to the paper \cite{IMT}, and also to  \cite{CoIv2}, \cite{CuIv}, \cite{CuIv2}.

\section{Dirichlet-Neumann operators and the Hamiltonian formulation of the governing equations}\label{sect4}

We introduce the Dirichlet-Neumann (DN) operators for the two fluid domains as follows. 
The outward normal for the domain $\Omega$ is denoted by ${\bf n}_s = (-\eta_x, 1)/\sqrt{1+\eta_x^2}.$ For the upper fluid domain $\Omega_1,$ 
the DN operator $G_1(\eta)$ is defined as
\begin{equation} \label{G1def}
    G_1(\eta)  (\varphi_{1})_s =-(\nabla\varphi_1)_s\cdot {\bf n}_s\sqrt{1+\eta_x^2}=-(\varphi_{1,z})_s +(\varphi_{1,x})_s\eta_x.
\end{equation} The minus sign is because the outward normal for the domain $\Omega_1$ is $- {\bf n}_s.$
For the lower fluid domain $\Omega$ the DN operator depends on the bottom variations, and is defined as 
 \begin{equation} \label{Gdef}
    G(\eta,\beta)  (\varphi)_s=(\nabla\varphi)_s\cdot {\bf n}_s\sqrt{1+\eta_x^2}=(\varphi_{z})_s -(\varphi_{x})_s\eta_x.
\end{equation}
There are recursive techniques for the computation of the DN operators, and for the most common configurations, the DN operators are computed in the form of perturbative expansions. These results will be used in the next section. 

However, it is important to point out that the governing equations \eqref{Ber} and \eqref{kin_interface}
for the motion of the pycnocline $z=\eta(x,t)$ can be written in a quasi-Hamiltonian form for the variables $\eta(x,t)$ and $\xi(x,t)=\rho (\varphi)_s - \rho_1 (\varphi_{1})_s .$  Such expression originates from the Hamiltonian approach in \cite{Z68} and its developments such as \cite{CraigGroves1, Craig1993,NearlyHamiltonian, CIMT,W, Curtin}  for a single layer and \cite{BB,Compelli1Wavemotion,CompelliIvanov1,CoIv2,CGK,CGS,CIM-16,CI-19,CuIv,CuIv2}, etc. for internal waves: 
\begin{equation}
\label{EOMsys2}
        \left\lbrace
        \begin{array}{lcl}
        \eta_t= \frac{\delta H}{\delta {\xi}}
        \\
        \xi_t=- \frac{\delta  H}{\delta \eta} -\Gamma  \partial_x ^{-1} \eta_t
        \end{array}
        \right.
\end{equation}
where $$\Gamma := \rho \gamma- \rho_1 \gamma_1,$$
with the functional $H(\xi,\eta),$  
\begin{multline}
\label{Main_Ham}
H(\eta,\xi)=\frac{1}{2}\int_{\mathbb{R}} \xi G(\eta, \beta) B^{-1}G_1(\eta)\xi \,dx
- \frac{1}{2}\rho\rho_1  (\gamma-\gamma_1)^2 \int _{\mathbb{R}} \eta \eta_x   B^{-1} (\eta \eta_x)  \,dx\\
-\int_{\mathbb{R}} (\gamma\eta+\kappa)\xi\eta_x \,dx+\rho_1\int _{\mathbb{R}}\mu B^{-1}G(\eta, \beta)\xi\,dx\\
+ \frac{\rho\gamma^2 - \rho_1 \gamma_1^2 }{6 } \int_{\mathbb{R}}\eta ^3 dx+\frac{g(\rho-\rho_1)+\kappa(\rho \gamma - \rho_1 \gamma_1)}{2} \int_{\mathbb{R}} \eta^2 dx,
\end{multline}
where $B:= \rho_{1}G+\rho G_{1}.$   
This expression formally coincides with the expression for the flat bottom \cite{CoIv2}, apart from the fact that $G(\eta,\beta)$ depends now on the bottom topography - see the derivation in \cite{IMT}.

Introducing the variable $\mathfrak{u}:=\xi_x$ (tangential "momentum") one can write down \eqref{EOMsys2} in the equivalent form
\begin{equation}
\label{EOMsys1}
        \left\lbrace
        \begin{array}{lcl}
        \eta_t=-\left(\frac{\delta H }{\delta {\mathfrak{u}}} \right)_x
        \\
        \mathfrak{u}_t+\Gamma \eta_t=-\left(\frac{\delta H}{\delta {\eta}} \right)_x .
        \end{array}
        \right.
\end{equation}
We note that due to the assumptions made, the variables $\eta, \xi $ and $\mathfrak{u}$ are in $\mathcal{S}(\mathbb{R})$ for all values of $t.$  

\section{Physical scales and approximations}\label{sect5}

We make the following assumptions about the relative scales of the physical quantities. We introduce the small-amplitude parametetr
$\varepsilon = \frac{|\eta|_{\text{max}}}{h}.$ The typical wavelengths $L$ are assumed significantly bigger than $h$, i.e. $$\delta=\frac{h}{L} \ll 1.$$ The wave number $k=2\pi/L $ is an eigenvalue or a Fourier multiplier for the operator $D=-i\partial_x$ (when acting on waves of the monochromatic waveforms $e^{ik(x-c(k) t)}$). Therefore $k h = 2 \pi \delta $  is of order $\delta.$ 
The further assumptions about the scales are

1. $\mathcal{O}(\delta)=\mathcal{O}(\varepsilon);$

2. The physical constants $h,$ $\rho,$ $\rho_1,$ $\gamma,$  $\gamma_1$ are $\mathcal{O}(1) .$ The rationale of this assumption is that we can work with non-dimensional quantities, such that $g=h=\rho=1,$ see for example \cite{CJ2}. Then all other constants and variables are replaced by non-dimensional combinations involving $g,\rho$ and $h$. However, in order to keep track of the quantities in the expressions, we will not write explicitly that the non-dimensional $g,\rho,$ and $h$ are equal to 1, but nevertheless the non-dimensionality of the quantities will be assumed. Then, for example, the non-dimensional $\eta $ is a quantity of order $\delta.$

3. $h_1 k=\mathcal{O}(1)$ and $h k=\mathcal{O}(\delta),$ i.e. $ h/h_1 = \mathcal{O}(\delta) \ll 1$. This corresponds to a deep upper layer. Since the operator $D$ has an eigenvalue $k$, we also keep in mind that $h_1 D=\mathcal{O}(1) $  and $h D=\mathcal{O}(\delta) .$  

4. $\xi=\mathcal{O}(1)$. Then, with the assumptions for the operator $\partial_x$ we have $\mathfrak{u}:=\xi_x=\mathcal{O}(\delta) .$

5. $|\beta|_{\text{max}}\ll \varepsilon^{1/3}$ and $\beta ' (x) = \mathcal{O}(\varepsilon).$  This is related to the size and the steepness of the bottom variations.

\newpage

The DN operator $G(\beta, \eta)$ is derived in \cite{IMT} and the result is
\footnote{The action of such operators $\mathcal{G}(D)$ which are expressed via $D,$ on functions $\mathfrak{f}(x)=\frac{1}{2\pi}\int e^{ikx} \hat{\mathfrak{f}}(k) dk $ is given by the integral $$ \mathcal{G}(D) \mathfrak{f} =\frac{1}{2\pi}
\int e^{ikx} \mathcal{G}(k)\hat{\mathfrak{f}}(k) dk . $$}
\begin{equation}
G(b, \eta)=\delta^2 D[ b(X)+\varepsilon \eta]D -\delta^4 D^2 \left[ \frac{1}{3}b^3(X)+\varepsilon h^2 \eta \right]D^2 + \delta^6 \frac{2}{15}h^5 D^6+\mathcal{O}(\delta^8, \varepsilon \delta^6, \varepsilon^2 \delta^4) , \label{DN_1}
\end{equation}
\noindent where $b(X):= h-\beta(\varepsilon x)=\mathcal{O}(1)$ is the local depth at location $x$ and the introduction of a "slow" variable $X=\varepsilon x$ indicates that the bottom depth varies slowly with $x,$ in a sense that 
\begin{equation}\label{smallDx}
    \partial_x b(X)= \mathcal{O}(\varepsilon),
    \end{equation}
while $b'(X)= \mathcal{O}(1),$ in accordance with the 5-th assumption above. The scale of each term in the expansion in \eqref{DN_1} is written in front of it. Applying the 1-st assumption, we obtain
\begin{equation}
G(b, \eta)=\delta^2 D b(X)D +\delta^3 D \eta D -\delta^4  \frac{1}{3} D^2 b^3(X) D^2 +\mathcal{O}(\delta^5). \label{DN_G}
\end{equation}
It is important to point out that from Assumption 5 it follows that the commutator between $b(X)$ and $D$ gives a quantity of smaller order, $(Db - bD)\mathfrak{f}= -i(b\mathfrak{f})_x-b(-i\mathfrak{f}_x)=-ib_x \mathfrak{f} = \varepsilon (-i b'(X)\mathfrak{f}).$

The assumption defining the scale for the intermediate long waves $0.05 < h_1/L < 0.5$ implies $\tanh(h_{1} {D})=\mathcal{O}(\tanh(h_{1} k))=\mathcal{O}(1).$
Then the DN operator $G_1$ has the expansion (see, for example, \cite{CGK})
\begin{equation} \label{DN_G1}
 {G}_{1}(\eta)=\delta D \tanh(h_{1} {D})-\delta^{3} D \eta D +\delta^{3} D \tanh(h_{1}{D})\eta {D}\tanh(h_{1}{D}) +\mathcal{O}(\delta^{5}).
\end{equation}
We need also the operator:
\begin{align} \label{E}
     GB^{-1}G_{1} & = G (\rho_{1}G+\rho G_{1})^{-1} G_{1}=\frac{1}{\rho}G\left[\left(1+\frac{\rho_{1}}{\rho}GG^{-1}_{1}\right)G_{1}\right]^{-1}G_{1}  \nonumber \\
     & =   \frac{1}{\rho}GG_{1}^{-1}\Big[1-\frac{\rho_{1}}{\rho}GG_{1}^{-1}+\frac{\rho_{1}^{2}}{\rho^{2}}(GG_{1}^{-1})^{2}-\cdots\Big]G_{1} {} \nonumber\\
     & =  \frac{1}{\rho}G-\frac{\rho_{1}}{\rho^2}GG_{1}^{-1}G+\frac{\rho_{1}^{2}}{\rho^{3}}GG^{-1}_{1}GG_{1}^{-1}G-\cdots .
 \end{align}
Since in the leading order 
\begin{equation}
    \left|\frac{\rho_{1}}{\rho}GG^{-1}_{1}\right|= \frac{\rho_{1}}{\rho} \left| \frac{\tanh(k h)}{\tanh(kh_1)}\right| \ll 1
    \end{equation}
as far as $|\tanh(k h)| \ll |\tanh(kh_1)|= \mathcal{O}(1),$ the expansion \eqref{E} is valid at least in some neigbourhood of $k=0.$
From the expansion \eqref{E} it is evident that the operator $GB^{-1}G_{1}$ is self-adjoint, since the operators $G$ and $G_1$ are self-adjoint, see, for example, \cite{CGK}. Therefore $GB^{-1}G_{1}=G_1B^{-1}G. $ With the expressions \eqref{DN_G} and \eqref{DN_G1} in \eqref{E} we obtain
\begin{align}
    GB^{-1}G_{1} & =  \delta^{2}\frac{1}{\rho}DbD+\delta^{3}\frac{1}{\rho}  D\eta D- \delta^{3}\frac{i\rho_1}{\rho^2} DbD\mathcal{T} bD \nonumber \\
    & - \delta^{4}\frac{i \rho_{1}}{\rho^{2}}\Big(D b \mathcal{T}D\eta D +D\eta D\mathcal{T} b D \Big) 
    - \delta^{4}\frac{1}{\rho} \Big(\frac{\rho_{1}^{2}}{\rho^{2}} D b D \mathcal{T} b \mathcal{T} D b D +\frac{1}{3}D^2 b ^3D^2\Big)+\mathcal{O}(\delta^{5}), \nonumber \\
\end{align}
where the operator $\mathcal{T} :=-i\coth(h_{1}D)$ is introduced. In a similar manner we can expand the other operators involving $B^{-1}.$ As a final result, from \eqref{Main_Ham} we obtain 
\begin{align}
         H(\eta,\mathfrak{u}) & =  \delta \frac{1}{2\rho}\int_{\mathbb{R}} b \mathfrak{u}^{2} dx+\delta\kappa\int_{\mathbb{R}}\eta\mathfrak{u}dx+\delta\frac{(\rho\gamma-\rho_{1}\gamma_{1})\kappa+g(\rho-\rho_{1})}{2}\int_{\mathbb{R}}\eta^{2} dx {}\nonumber\\
    &- 
    \delta^{2}\frac{\rho_{1}}{2\rho^{2}}\int_{\mathbb{R}} b\mathfrak{u}  \mathcal{T}(b\mathfrak{u})_{x} dx+\delta^{2}\frac{\gamma}{2}\int_{\mathbb{R}}\eta^{2}\mathfrak{u} dx+\delta^{2}\frac{1}{2\rho}\int_{\mathbb{R}}\eta\mathfrak{u}^{2} dx         +   \delta^{2}\frac{\rho\gamma^{2}-\rho_{1}\gamma_{1}^{2}}{6}\int_{\mathbb{R}}\eta^{3} dx    \nonumber\\
    &+ 
   \delta^{3}\frac{1}{2\rho}\int_{\mathbb{R}}\Big(\frac{\rho_{1}^{2}}{\rho^{2}} b [\mathcal{T}(b \mathfrak{u})_{x}]^2-\frac{b^3}{3}\mathfrak{u}_{x}^2\Big) dx {}\nonumber\\
    &- 
    \delta^{3}\frac{\rho_{1}}{\rho^{2}}\int_{\mathbb{R}} \eta\mathfrak{u}\mathcal{T} (b\mathfrak{u})_{x} dx-\delta^{3}\frac{\rho_{1}(\gamma-\gamma_{1})}{2\rho}\int_{\mathbb{R}}\eta^{2}\mathcal{T}(b\mathfrak{u})_{x}dx . 
    \label{Ham}
\end{align}
The equations obtained from \eqref{EOMsys1} with \eqref{Ham} are
\begin{align}
    \eta_{t}+\kappa\eta_{x}&+\frac{1}{\rho}(b\mathfrak{u})_{x}+\delta\frac{1}{\rho}(\eta\mathfrak{u})_{x}
    -\delta\frac{\rho_{1}}{\rho^{2}}(b\mathcal{T} (b \mathfrak{u})_x  )_x +
    \delta\gamma\eta\eta_{x}+\delta^{2}\frac{b^{3}}{\rho}\Big(\frac{\rho_{1}^{2}}{\rho^{2}}\mathcal{T}^{2}+\frac{1}{3}\Big)\mathfrak{u}_{xxx}{}\nonumber \\
    &-\delta^{2}\frac{\rho_{1}b}{\rho^{2}}((\eta\mathcal{T}\mathfrak{u}_{x})_{x}+\mathcal{T}(\eta\mathfrak{u})_{xx})-\delta^{2}\frac{\rho_{1}b(\gamma-\gamma_{1})}{2\rho}\mathcal{T}(\eta^{2})_{xx}=\mathcal{O}(\delta^3), \label{eta-t}
\end{align}
\begin{align}
    \mathfrak{u}_{t}+\kappa\mathfrak{u}_{x}&+\Gamma\eta_{t}+A\eta_{x}+\delta\frac{1}{\rho}\mathfrak{u}\mathfrak{u}_{x}+\delta\gamma(\eta\mathfrak{u})_{x}+\delta(\rho\gamma^{2}-\rho_{1}\gamma_{1}^{2})\eta\eta_{x}{}\nonumber \\
   &-\delta^{2}\frac{\rho_{1}b}{\rho^{2}}(\mathfrak{u}\mathcal{T}\mathfrak{u}_{x})_{x}-\delta^{2}\frac{\rho_{1}(\gamma-\gamma_{1})b}{\rho}(\eta\mathcal{T}\mathfrak{u}_{x})_{x}=\mathcal{O}(\delta^3) ,\label{u-t}
\end{align}
where $A=(\rho\gamma-\rho_{1}\gamma_{1})\kappa+g(\rho-\rho_{1})=g(\rho-\rho_{1})+\Gamma \kappa.$  Since the commutator between $b(X)$ and $D$ gives a quantity of smaller order, in the terms of order $\delta^2$ the local depth $b(X)$ is written before the differential operators  \footnote{Indeed, any change in the order of $b(X)$ and $D$ will result in the appearance of the commutator. However, $\delta^2 D b(X)= \delta ^2  b(X)D + \delta^2 [D, b(X)]= \delta^2  b(X)D - \delta^2  \delta i b’(X)=  \delta^2   b(X)D+ \mathcal{O}(\delta^3)$. The $ \mathcal{O}(\delta^3)$ - terms are being neglected. }.

The obtained system describes intermediate long internal waves. The intermediate waves (or transitional
waves) for this internal wave system are the waves with $ 0.05 < (h_1/L) < 0.5$. Then $0.30=\tanh (2\pi \cdot 0.05)<\tanh (k h_1)< \tanh (2\pi \cdot 0.5)=0.996 $ and the assumption $\tanh(h_{1} {D})=\mathcal{O}(1)$ is justified. Moreover, when the ratio $h_1/L$
is close to 0.5, then $|\tanh(kh_1)|\approx 1$ and hence we can make the approximation $\tanh(h_1 D) = \text{sign}(D). $
Accordingly $$\mathcal{T}= -i\coth(h_1D) \to -i \,\text{sign}(D)= \mathcal{H},$$ the Hilbert transform, and $\qquad \mathcal{T} \partial_x \to |D|. $  

\section{The Intermediate Long Wave equation} \label{sect6}

The terms of the leading order in \eqref{eta-t} and \eqref{u-t} satisfy the following system of linear equations with ``slowly varying''  coefficients:
\begin{equation} \label{lead-o}
\left\{
\begin{array}{ll}
     & \eta_{t}+\kappa \eta_x +\frac{b(X)}{\rho}\mathfrak{u}_{x}=0, \\
     & \mathfrak{u}_{t}+\kappa \mathfrak{u}_{x}+ \Gamma\eta_{t}+A\eta_{x}=0.\\
\end{array}\right.
\end{equation}
Thus we look for a monochromatic solution for $\eta$ and $\mathfrak{u}$ in the form
\begin{equation}\label{los}
\begin{split}
        \eta(x,t)&=\eta_0e^{ik(x-c(X)t)}\\
\mathfrak{u}(x,t)&=\mathfrak{u}_0e^{ik(x-c(X)t)}
\end{split}
\end{equation}
where $c(X)$ is the wave speed, which depends on the ``slowly varying'' variable $X$. We obtain a quadratic equation for $c(X):$ 
\begin{equation}\label{eq4c}
    (c-\kappa)^2+ \frac{b(X)}{\rho}\Gamma (c-\kappa)-\frac{g b(X)(\rho-\rho_1)}{\rho}=0,
\end{equation} giving two solutions for the left- and for the right-travelling waves (when the plus and the minus signs are taken respectively):
\begin{equation}\label{c}
    c(X)= \kappa- \frac{\Gamma b(X)}{2\rho} \pm \sqrt{ \frac{\Gamma^2 b^2(X)}{4\rho^2} + \frac{g b(X)(\rho-\rho_1)}{\rho}}.
\end{equation}
From any of the equations in \eqref{lead-o} with \eqref{los} we also obtain in the leading order
\begin{equation}\label{u-lead}
     \mathfrak{u}=\frac{\rho(c(X)-\kappa) }{b(X)} \eta.
\end{equation}
Let us now consider the problem \eqref{eta-t} - \eqref{u-t} in the next order, that is, neglecting the $\delta^2$- terms. The system of equations reduces to
\begin{align}
    \eta_{t}+\kappa\eta_{x}&+\frac{1}{\rho}(b\mathfrak{u})_{x}+\delta\frac{1}{\rho}(\eta\mathfrak{u})_{x}
    -\delta\frac{\rho_{1} b^2}{\rho^{2}}\mathcal{T}  \mathfrak{u}_{xx} +
    \delta\gamma\eta\eta_{x}= \mathcal{O}(\delta^2)  , \label{eta-t-1} \\
    \mathfrak{u}_{t}+\kappa\mathfrak{u}_{x}&+\Gamma(\eta_{t} +\kappa\eta_{x})+g(\rho-\rho_{1})\eta_{x}+\delta\frac{1}{\rho}\mathfrak{u}\mathfrak{u}_{x}+\delta\gamma(\eta\mathfrak{u})_{x}+\delta(\rho\gamma^{2}-\rho_{1}\gamma_{1}^{2})\eta\eta_{x}   
   =\mathcal{O}(\delta^2) .\label{u-t-1}
\end{align}
Introducing the notation $c_0(X)=c(X)-\kappa,$ using \eqref{u-lead} and the structure of the terms in \eqref{eta-t-1} - \eqref{u-t-1}, we are looking for a relation between $ \mathfrak{u}$ and $\eta$ involving $\mathcal{O}(\delta)$ approximations:
\begin{equation}\label{u-eta}
     \mathfrak{u}=\frac{\rho c_0(X) }{b(X)} \eta+ \delta f(X) \partial_x ^{-1}\eta + \delta \alpha_{1}\eta^{2}+\delta\alpha_{2}\mathcal{T} \eta_{x}+\mathcal{O}(\delta^2),
\end{equation}
for some yet undetermined constants $\alpha_{1}$ and $\alpha_{2}$ and slowly-varying function $f(X).$  
Differentiating \eqref{eq4c} with respect to $X$ we obtain 
\begin{equation}\label{c0prim-eq}
    c_0'(X)=\frac{\rho c_0^2 }{b(2\rho c_0+b\Gamma)}b'(X).
\end{equation}
The compatibility of the two equations \eqref{eta-t-1} - \eqref{u-t-1} under \eqref{u-eta} determines uniquely the unknown quantities as 
\begin{align}
    f(X)&=- \frac{\rho c(\rho c_0 + \Gamma b)}{b c_0(2\rho c_0 + \Gamma b) } c_0' (X) ,\label{f} \\ 
    \alpha_1&=\frac{\rho[\gamma \rho b c_0 + b^2(\rho \gamma^2 -\rho_1 \gamma_1^2)-\rho c_0^2 - 2\Gamma b c_0-\gamma \Gamma b^2]}{2b^2(2\rho c_0+ \Gamma b)}, \label{alpha-1} \\
    \alpha_2&= \frac{\rho_1 c_0 (\rho c_0+ \Gamma b)}{2\rho c_0+ \Gamma b}. \label{alpha-2}
\end{align}
The resulting equation for the elevation $\eta$ is
\begin{align}\label{eq4eta}
    \eta_t + c(X) \eta_x +\delta \frac{\rho c_0^2-\kappa(\rho c_0+\Gamma b)}{c_0(2\rho c_0+\Gamma b)}  c'_0 (X)  \eta  & -\delta  \frac{\rho_{1}bc_0^{2}}{2\rho c_0 +\Gamma b}  \mathcal{T} \eta_{xx}  \nonumber  \\
        &+ \delta \frac{3\rho c_0^{2} +3\gamma\rho b c_0+b^{2}(\rho\gamma^{2}-\rho_{1}\gamma_{1}^{2})}{b(2 \rho c_0 +\Gamma b)} \eta\eta_{x}=\mathcal{O}(\delta^2) .
\end{align}
This is an equation with slowly varying coefficients, depending on the local depth of the lower layer $b(X).$ We observe that the  formation of nonlinear waves is a balance between three effects of the same magnitude, described by the three terms of order $\delta$: bottom variations, dispersion and nonlinearity.

A special situation is possible when the coefficient of the nonlinear term is zero, 
\begin{equation}\label{crit}
    3\rho c_0^{2} +3\gamma\rho b c_0+b^{2}(\rho\gamma^{2}-\rho_{1}\gamma_{1}^{2}) =0.
\end{equation}
This condition, together with \eqref{eq4c} can be satisfied for a certain depth $b= b^*>0$, if  
\begin{equation}
\label{eq78}
   \gamma_{1}^2 >\frac{\rho}{4\rho_{1}} \gamma^2.
\end{equation}
In particular, when $\gamma=0,$ (and $U(z)=0,$ $\kappa=0,$ that is, when there is no shear current in the lower layer) the condition \eqref{crit} has a solution $$c_0^*=\pm \sqrt{\frac{\rho_1}{3\rho}}  \gamma_1 b,$$
and from \eqref{eq4c} we obtain the critical depth
\begin{equation}
    b^*= \frac{g(\rho-\rho_1)}{\rho_1 \gamma_1^2\left( \frac{1}{3}+ \sqrt{\frac{\rho_1}{3\rho}} \right)}
\end{equation} for the minus sign of $c_0^*,$ otherwise $b^*$ will be negative. Given the fact that for the Equatorial Undercurrent usually $\gamma_1<0,$ then the critical depth would appear for the eastward running wave  
$c_0^*=- \sqrt{\frac{\rho_1}{3\rho}}  \gamma_1 b >0.$

When the nonlinear term is missing, the remaining terms in \eqref{eq4eta} are linear, and in order to analyse the solitary waves one needs to consider the higher order terms. In the case of constant depth, $b=h$ = constant, these are provided in Appendix 1 as an illustration. We note that in the absence of shear currents the nonlinear term $\eta \eta_x$ is always present.  

In the case of constant depth, the equation \eqref{eq4eta} is integrable, it is known as Intermediate Long Wave Equation (ILWE). 
The soliton theory for ILWE has been developed in a number of works \cite{ChLee,JoEg,Kod,Ma2,SaAbKo}, see also the details in the Appendix in \cite{CuIv}. In this case we can rewrite equation \eqref{eq4eta} in the form of an equation with constant coefficients (without the scaling)
\begin{equation}
\label{ILWEcc}
\eta_{t}+(c_0+\kappa)\eta_{x}+\mathcal{A}_{1}\eta\eta_{x}-\mathcal{A}_{2}\mathcal{T} \eta_{xx}=0,
\end{equation}
where
\begin{equation}\label{A12}
    \mathcal{A}_{1}:=\frac{3c_0^{2}\rho+3\gamma\rho c_0h+h^{2}(\rho\gamma^{2}-\rho_{1}\gamma_{1}^{2})}{h(2c_0\rho+\Gamma h)} , \ \mathcal{A}_{2}:=\frac{\rho_{1}hc_0^{2}}{2c_0\rho+\Gamma h}.
\end{equation}
We mention that \eqref{ILWEcc} has a one-soliton solution of the form
\begin{equation}
\label{eq72}
    \eta(x,t)=\frac{2\mathcal{A}_{2}}{\mathcal{A}_{1}}\cdot\frac{k_{0}\sin(k_{0}h_{1})}{\cos(k_{0}h_{1})+\cosh[k_{0}(x-x_{0}-(c_0+\kappa -\mathcal{A}_{2}k_{0}\cot(k_{0}h_{1}))t)]},
\end{equation}
where $0<k_{0}<\frac{\pi}{h_{1}}$. In the above formula $x_{0}$ and $k_{0}$ are the soliton parameters, i.e. arbitrary constants within their range of allowed values, $x_{0}$ is the initial position of the crest of the soliton and $k_{0}$ is related to its amplitude.

\section{ILWE limits to other models}\label{sect7}

In the short-wave limit (or, "deep water" limit, e.g. when the ratio $h_1/L \ge 0.5$) the operator $ \mathcal{T}$ becomes the Hilbert transform $ \mathcal{H}$ and the equation \eqref{ILWEcc} becomes the well known Benjamin-Ono (BO) equation \cite{BO1,BO2}, see also \cite{CI19,CuIv}.
Like the ILWE, the BO is an integrable equation whose solutions can be obtained by the Inverse Scattering method \cite{FA,KM98,Ma1}.

The one-soliton solution can be written in the form
\begin{equation} \eta(x,t)=\frac{\eta_0}{1+\left(\frac{\eta_0 \mathcal{A}_1}{4\mathcal{A}_2}\right)^2[x-x_0-(c+\frac{1}{4}  \mathcal{A}_1\eta_0) t]^2} ,\end{equation} where the constants are the initial soliton crest position $x_0$ and the soliton amplitude $\eta_0.$  The one-soliton solution can be obtained as a limit from \eqref{eq72}, see for example \cite{CuIv}.

On the other hand, when $h_1/L$ is close to 0.05, $kh_1=2\pi \cdot 0.05=0.314$ and we can expand 
\begin{equation}
    \mathcal{T}=-i \coth(h_1D)=-i \left((h_1D)^{-1}+\frac{1}{3} h_1D -\frac{1}{45}(h_1D)^3+\ldots \right),
\end{equation} 
so that \eqref{ILWEcc} leads to the integrable Korteweg - de Vries \cite{KdV} equation (or KdV for short)
\begin{equation}
\label{KdV}
\eta_{t}+\left(c_0+\kappa-\frac{\mathcal{A}_{2}}{h_1}\right)\eta_{x}+\mathcal{A}_{1}\eta\eta_{x}+ \frac{h_1}{3}\mathcal{A}_{2} \eta_{xxx}=0.
\end{equation}
This KdV equation differs from the KdV model (for long waves, e. g. $h_1/L\le 0.05$) obtained in \cite{Iv17} in a similar setting, where the surface waves are assumed of small amplitude in comparison to the internal waves, however under the different assumption, $\varepsilon = \mathcal{O}(\delta^2).$  Nevertheless, the expression for $c_0,$ the wave propagation speed in the leading order for the model in \cite{Iv17} is the same. The inverse scattering method for the KdV equation is presented, for example, in \cite{ZMNP}. 

\section{Variable bottom and adiabatic invariants}\label{sect8}

In the case of slowly variable bottom the coefficients of the model equation vary slowly with $x$ and therefore the conserved quantities of the
integrable ILWE equation (see, for example, \cite{Kod}) provide some analogues which can be used for qualitative estimates. These quantities change very slowly over time and are also known as adiabatic invariants. Let us make the simplifying assumptions $\gamma=\gamma_1=0,$ $\kappa=\kappa_1=0,$ and introduce the variable $E(x,t)=\sqrt{c(X)} \eta(x,t)$ where from \eqref{eq4c}\begin{equation}
    c(X)=\sqrt{\frac{g(\rho-\rho_1)b(X)}{\rho}},
\end{equation} and from \eqref{eq4eta}
\begin{align}\label{eq4E}
    \left(\frac{1}{c(X)}E\right)_t + E_x & -\delta  \frac{\rho_{1}b}{2\rho}  \mathcal{T} E_{xx}  + \delta \frac{3  }{2 b  \sqrt{c} } E E_{x}=\mathcal{O}(\delta^2) ,
\end{align} which with \eqref{smallDx} can be written also as
\begin{align}\label{eq4E2}
    \left(\frac{1}{c(X)}E\right)_t + \left(E  -\delta  \frac{\rho_{1}b}{2\rho}  \mathcal{T} E_{x}  + \delta \frac{3  }{ 4b  \sqrt{c} } E^2 \right)_x =\mathcal{O}(\delta^2) .
\end{align}

\begin{figure}[!ht]
\centering
\includegraphics[width=0.5 \textwidth]{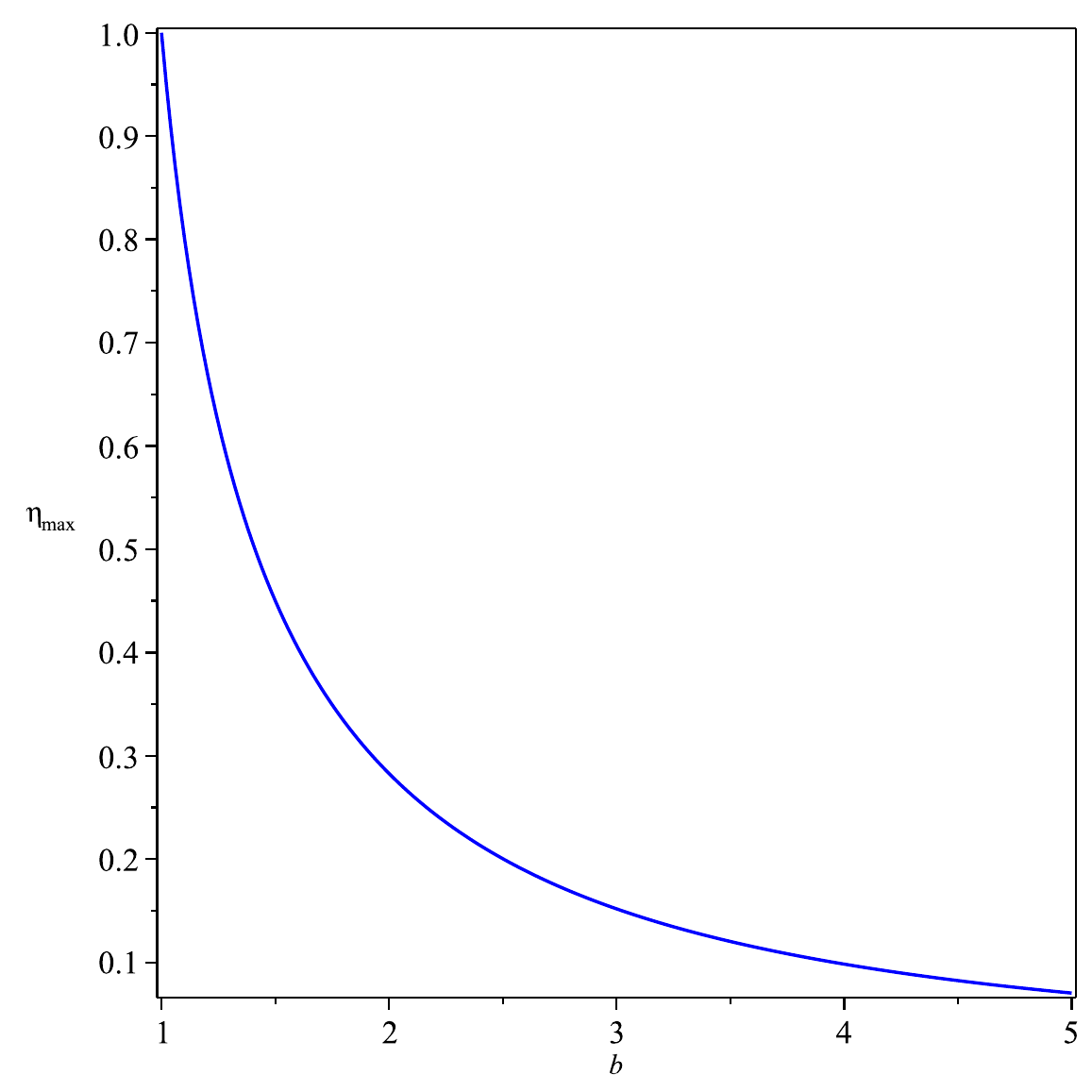}
 \caption{The dependence \eqref{etaMaxb} with $\mu_1=1,$ $\mu_2=\pi/4$.}\label{FigEta}
\end{figure}

Like in the case with constant coefficients, there are conserved quantities such as the soliton mass, $\mathcal{M}$ and the soliton energy, $\mathcal{E},$ 
\begin{align}
\mathcal{M} &=  \int_{-\infty}  ^{\infty}  \frac{E(x,t)}{c(X)} dx ,              \\ 
\mathcal{E} &= \frac{1}{2}\int_{-\infty}  ^{\infty}  \frac{E^2(x,t)}{c(X)} dx,
    \end{align}
etc., which are now the adiabatic invariants that we use in order to estimate the influence of the variable bottom. The "approximate" one-soliton solution of \eqref{eq4E} in the case of slowly varying coefficients is just the one-soliton solution of the ILWE with constant coefficients described in the previous section, with constants, replaced as follows: $h\ \to b(X)$ ,  $c_0+\kappa \to c(X):$
\begin{align}
        E(x,t)&=\frac{2 \rho_1 b^2 \sqrt{c}}{3 \rho}\cdot \frac{k_0 \sin(k_0 h_1)}{\cos(k_0 h_1)+\cosh[k_{0}(x-x_{0})-k_0(c -\delta\mathcal{A}_{2}k_{0}\cot(k_{0}h_{1}))t]}, \\
        &\text{where} \, \, \mathcal{A}_{2} =\frac{\rho_1 bc}{2\rho}, \quad 0<k_0h_1 < \pi. \nonumber
\end{align}
With the integral formula \eqref{I1}, neglecting the $x$-dependence of the slowly varying variables in comparison to the fast growing cosh function,   we obtain
\begin{equation}
    \mathcal{M}\approx \frac{4\rho_1 b^2}{3\rho  \sqrt{c}} (k_0 h_1),
\end{equation} and therefore 
\begin{equation}
    k_0 \approx \frac{3\rho  \mathcal{M}  \sqrt{c}}{4\rho_1 h_1 b^2}<\frac{\pi}{h_1}.
\end{equation}
The last formula suggests that instead of the soliton parameter $k_0$ it is better to use the adiabatic invariant $\mathcal{M}$ as a constant parameter during the motion.
The amplitude of the soliton wave is therefore
\begin{align}
    \eta_{max}&=\frac{E_{max}}{\sqrt{c}}=\frac{2 \rho_1 b^2 k_0 \sin(k_0 h_1) }{3 \rho(\cos(k_0 h_1)+1)}= \frac{2 \rho_1 b^2 k_0  }{3 \rho}\tan \frac{k_0 h_1}{2} \nonumber \\
    &\approx  \frac{\mathcal{M}\sqrt{c}}{2 h_1 }\tan \frac{3\rho \mathcal{M}\sqrt{c}}{8\rho_1 b^2 } 
    = \frac{\mathcal{M}}{2 h_1 }\left(\frac{g(\rho-\rho_1)}{\rho}\right)^{1/4} b^{1/4}\tan \left [\frac{3\rho \mathcal{M}}{8\rho_1  } \left(\frac{g(\rho-\rho_1)}{\rho}\right)^{1/4} b^{-7/4} \right ],
   \end{align}
and as a function of the depth $b$ of the lower layer, we have the approximation
\begin{equation}\label{etaMaxb}
    \eta_{max}(b)\approx {\mu_1} b^{1/4} \tan(\mu_2 b^{-7/4})
    \end{equation} with the constant parameters (adiabatic invariants)
    \begin{equation}
        \mu_1= \frac{\mathcal{M}}{2 h_1 }\left(\frac{g(\rho-\rho_1)}{\rho}\right)^{1/4} , \quad \mu_2 = \frac{3\rho \mathcal{M}}{8\rho_1  } \left(\frac{g(\rho-\rho_1)}{\rho}\right)^{1/4}  .
    \end{equation}
The dependence \eqref{etaMaxb} is illustrated in Fig. \ref{FigEta}. It shows, as expected from the shallow-water theory, that the soliton amplitude grows rapidly when the depth of the lower layer (hence, the overall depth) decreases. This illustration is only a rough estimate as a result of all the approximations made. We note that it was implicitly assumed, in addition, that there is no birth of new solitons, which is a typical process, when the soliton moves from a deep to a shallow lower layer. In the long-wave regime the process is studied in \cite{DR,IMT}.    

Perhaps a more accurate estimate for $\eta_{max}(b)$ can be obtained from the energy consideration. In the long-wave regime this is reported, for example, in \cite{Tal}.

\section{Conclusions}

We have presented a study of the propagation of interfacial internal waves in the case when the lower layer is significantly thinner than the upper layer. The flat surface approximation is assumed. In addition, we include the effects of current with a specific profile as well as variable bottom. In a sense, this situation completes the analysis of possible two-layer configurations-the one with a thin upper layer, presented in \cite{CuIv} and the case where the two layers have depths of similar magnitudes \cite{CoIv2,IMT,HIS}. The situation with a thin lower layer is less common in practice, it corresponds to a deep thermocline,  nevertheless, it is quite realistic, given the seasonal variations of the thermocline, pointed out in the introduction. 
We obtained a scalar model equation, known as the Intermediate Long Wave Equation (ILWE), which is integrable in the case of constant depth. In the case of variable bottom, soliton solutions most likely exhibit the effects of breaking, such as those in the case of the KdV equation (with variable coefficients) reported in \cite{IMT}. These effects deserve further investigation. We identified a critical depth of the lower layer (which depends on the current) when the nonlinear term $\eta \eta_x$ of the ILWE vanishes. This effect is not possible in the absence of a shear current. We have also identified the nonlinearities in the higher-order approximation which play a role in the absence of the $\eta \eta_x$ term - these are the terms with $\eta^2 \eta_x$ and the nonlocal terms $\eta_{x}\mathcal{T}\eta_{x}$ and $\mathcal{T}(\eta^{2})_{xx}.$  Such nonlocalities do not appear in the higher-order models in the long-wave regime \cite{HIS}. The higher-order ILWE is not integrable and needs further analysis.

\subsection*{Data availability statement }

The reported results are of purely theoretical nature, and no data have been used to support these results.

\subsection*{Acknowledgments} This publication has emanated from research conducted with the financial support of Taighde \' Eireann – Research Ireland under Grant number 21/FFP-A/9150. The authors thank an anonymous referee whose suggestions helped to significantly improve the publication. 

\section*{Appendix 1}  

Here we provide the one-component equation for $\eta$ which is equivalent to the second order system \eqref{eta-t} - \eqref{u-t} with the simplifying assumption of flat bottom $b=h=\text{const.}$  The
relation \eqref{u-eta} is extended with terms of order $\delta^{2} $ as follows:
\begin{align}
    \label{eq113}
    \mathfrak{u}=\frac{\rho}{h}c_0\eta&+\delta\alpha_{1}\eta^{2}+\delta\alpha_{2}\mathcal{T} \eta_{x}+\delta^{2}\alpha_{3}\eta_{xx}+\delta^{2}\alpha'_{3}\mathcal{T}^{2}\eta_{xx}  \nonumber \\
    &+\delta^{2}\alpha_{4}\eta\mathcal{T} \eta_{x}+\delta^{2}\alpha'_{4}\partial^{-1}(\eta_{x}\mathcal{T} \eta_{x})+\delta^{2}\alpha_{5}\mathcal{T} (\eta^{2})_{x}+\delta^{2}\alpha_{6}\eta^{3} .
    \end{align}
The constants $\alpha_1$ and $\alpha_2$ are as in \eqref{alpha-1} - \eqref{alpha-2} with $b=h.$ The new unknown constants are $\alpha_{3}$, $\alpha'_{3}$, $\alpha_{4}$, $\alpha'_{4}$, $\alpha_{5}$ and $\alpha_{6}.$  The substitution of \eqref{eq113} in  \eqref{eta-t} - \eqref{u-t} and the comparison of all terms leads to the following expressions
\begin{equation}
    \alpha_{3}=-\frac{\rho hc_0(\rho c_0+\Gamma h)}{3(2\rho c_0+\Gamma h)},
\end{equation}
\begin{equation}
    \alpha'_{3}=-\frac{\rho_{1}^{2}hc_0^{2}(\rho c_0+\Gamma h)^{2}}{(2\rho c_0+\Gamma h)^{3}},
\end{equation}
\begin{equation}
    \alpha_{4}=\frac{\rho\rho_{1}c_0\Big[h^{2}c_0(\Gamma-\rho\gamma)^{2}-\rho c_0(\rho c_0^{2}+\rho\gamma hc_0+\rho_{1}\gamma_{1}^{2}h^{2})+\gamma_{1}h(2\rho c_0+\Gamma h)^{2}\Big]}{h(2\rho c_0+\Gamma h)^{3}},
\end{equation}
\begin{equation}
    \alpha'_{4}=\frac{\rho^{2}\rho_{1}c_0^{2}\Big[\rho c_0^{2}-\rho\gamma hc_0+2\Gamma hc_0+\gamma\Gamma h^{2}-h^{2}(\rho\gamma^{2}-\rho_{1}\gamma_{1}^{2})\Big]}{h(2\rho c_0+\Gamma h)^{3}},
\end{equation}
\begin{equation}
    \alpha_{5}=\frac{\rho_{1}(\rho c_0+\Gamma h)\Big[(\rho c_0+\Gamma h)\Big(3\rho c_0(c_0+\gamma h)+h^{2}(\rho\gamma^{2}-\rho_{1}\gamma_{1}^{2})\Big)-\gamma_{1}h(2\rho c_0+\Gamma h)^{2}\Big]}{2h (2\rho c_0+\Gamma h)^{3}},
\end{equation}
\begin{align}
        \alpha_{6}=\frac{\rho }{6h^{3}(2\rho c_0+\Gamma h)^{3}} &  \Big[\rho c_0^{2}-\rho\gamma hc_0+2\Gamma hc_0+\gamma\Gamma h^{2}-h^{2}(\rho\gamma^{2}-\rho_{1}\gamma_{1}^{2})\Big] \nonumber \\
                & \times \Big[2\rho\Big(3c_0(\rho c_0+\Gamma h)+h^{2}(\rho\gamma^{2}-\rho_{1}\gamma_{1}^{2})\Big)+3\Gamma h^{2}(\Gamma-\rho\gamma)\Big] .
\end{align} The wave speed in the leading order is
\begin{equation}
    c_0=    - \frac{\Gamma h}{2\rho} \pm \sqrt{ \frac{\Gamma^2 h ^2}{4\rho^2} + \frac{g h(\rho-\rho_1)}{\rho}},
\end{equation} we note that there are left and right propagation speeds.
Then the equation for $\eta$ is
\begin{align}\label{HILWE}
        \eta_{t} & +  (c_0+\kappa)\eta_{x} +\delta\mathcal{A}_{1}\eta\eta_{x}-\delta\mathcal{A}_{2}\mathcal{T} \eta_{xx} + \delta^{2}\mathcal{A}_{3}\eta_{xxx}+\delta^{2}\mathcal{A}'_{3}\mathcal{T} ^{2}\eta_{xxx}  \nonumber \\ 
     &+ \delta^{2}\mathcal{A}_{4} (\eta\mathcal{T} \eta_{x})_{x} +\delta^{2}\mathcal{A}'_{4}\eta_{x}\mathcal{T}\eta_{x} + \delta^{2}\mathcal{A}_{5}\mathcal{T}(\eta^{2})_{xx} + \delta^{2} \mathcal{A}_{6}(\eta^{3})_{x}=\mathcal{O}(\delta^3),\end{align}
where $\mathcal{A}_{1}$ and $\mathcal{A}_{2}$ are as in \eqref{A12} and
\begin{equation}
    \mathcal{A}_{3}=\frac{\rho h^{2}c_0^{2}}{3(2\rho c_0+\Gamma h)},
\end{equation}
\begin{equation}
    \mathcal{A}'_{3}=\frac{\rho_{1}^{2}h^{2}c_0^{3}(3\rho c_0+2\Gamma h)}{(2\rho c_0+\Gamma h)^{3}},
\end{equation}
\begin{equation}
    \mathcal{A}_{4}=\frac{\rho_{1}c_0\Big[\gamma_{1}h(2\rho c_0+\Gamma h)^{2}+\rho c_0\Big(h^{2}(\rho\gamma^{2}-\rho_{1}\gamma_{1}^{2})-\gamma h(\rho c_0+2\Gamma h)-c_0(5\rho c_0+4\Gamma h)\Big)\Big]}{(2\rho c_0+\Gamma h)^{3}},
\end{equation}
\begin{equation}
    \mathcal{A}'_{4}=\frac{\rho\rho_{1}c_0^{2}\Big[\rho c_0^{2}-\rho\gamma hc_0+2\Gamma hc_0+\gamma\Gamma h^{2}-h^{2}(\rho\gamma^{2}-\rho_{1}\gamma_{1}^{2})\Big]}{(2\rho c_0+\Gamma h)^{3}},
\end{equation}
\begin{equation}
    \mathcal{A}_{5}=\frac{\rho_{1}c_0\Big[\gamma_{1}h(2\rho c_0+\Gamma h)^{2}-(3\rho c_0+2\Gamma h)\Big(3\rho c_0(c_0+\gamma h)+h^{2}(\rho\gamma^{2}-\rho_{1}\gamma_{1}^{2})\Big)\Big]}{2(2\rho c_0+\Gamma h)^{3}},
\end{equation}
\begin{align}
        \mathcal{A}_{6}=\frac{\rho}{6h^{2}(2\rho c_0+\Gamma h)^{3}}    &
    \Big[-\rho c_0^{2}+\rho\gamma hc_0-2\Gamma hc_0-\gamma\Gamma h^{2}+h^{2}(\rho\gamma^{2}-\rho_{1}\gamma_{1}^{2})\Big]
    \nonumber \\
   & \times \Big[2\Big(3c_0(\rho c_0+\Gamma h)-h^{2}(\rho\gamma^{2}-\rho_{1}\gamma_{1}^{2})\Big)+3\gamma\Gamma h^{2}\Big]
        .
\end{align}
The higher-order intermediate long wave equation \eqref{HILWE} is most likely not integrable. We observe that if, for some choice of parameters, $\mathcal{A}_1=0,$ the remaining nonlinearities are those in the second line of \eqref{HILWE}. Apart from the cubic term there are also the nonlocal nonlinearities $\eta_{x}\mathcal{T}\eta_{x}$ and $\mathcal{T}(\eta^{2})_{xx}.$

\section*{Appendix 2} 

In the computation of the adiabatic invariants, the following integrals are useful. They can be computed by standard techniques, or with mathematical software,
\begin{align}
\label{I1}
    \int_{-\infty}^{\infty} \frac{dx}{\cos{\mathbf{a}}+\cosh(x)}&= 2\frac{\mathbf{a}}{\sin{\mathbf{a}}}, \quad 0< \mathbf{a} < \pi, \\
      \int_{-\infty}^{\infty} \frac{dx}{(\cos{\mathbf{a}}+\cosh(x))^2}&=2\frac{\sin{\mathbf{a}}-\mathbf{a}\cos{\mathbf{a}}}{\sin^3{\mathbf{a}}},
      \quad 0< \mathbf{a} < \pi.
\end{align}
The second integral follows from the first one by differentiation with respect to the parameter $\mathbf{a}.$


\end{document}